\documentclass[aip,preprint]{revtex4-1}

\pdftrailerid{}
\usepackage[T1]{fontenc} 
\usepackage{mathtools}
\usepackage{graphicx}
\usepackage{lineno}
\usepackage{physics}
\usepackage{mathrsfs}
\usepackage{bm}
\usepackage{float}

\usepackage{hyperref}
\usepackage[section]{placeins}
\usepackage{longtable}
\usepackage{multirow}
\usepackage{booktabs}
\usepackage{makecell}
\usepackage{xcolor}
\usepackage{dcolumn}
\usepackage[version=4]{mhchem}
\usepackage{threeparttable}

\usepackage[
margin=0.8in,
]{geometry}

\makeatletter
\newcommand{\warninput}[1]{\filename@parse{#1}\InputIfFileExists{#1}{}{\message{LaTeX Warning: File `\filename@base.\ifx\filename@ext\relax tex\else\filename@ext\fi' not found on input line \the\inputlineno}}}
\makeatother

\begin{document}
    \title{\textbf{Impact of scissors-correction schemes on first-principles calculations of second-harmonic generation in ultraviolet nonlinear-optical crystals}}

    \author{YingXing Cheng}
    \affiliation{Research Center for Crystal Materials; CAS Key Laboratory of Functional Materials and Devices for Special Environmental Conditions; Xinjiang Key Laboratory of Functional Crystal Materials; Xinjiang Technical Institute of Physics and Chemistry, Chinese Academy of Sciences, 40-1 South Beijing Road, Urumqi 830011, China.}
    \affiliation{Center of Materials Science and Optoelectronics Engineering, University of Chinese Academy of Sciences, Beijing 100049, China.}

    \author{Congwei Xie}
    \affiliation{Research Center for Crystal Materials; CAS Key Laboratory of Functional Materials and Devices for Special Environmental Conditions; Xinjiang Key Laboratory of Functional Crystal Materials; Xinjiang Technical Institute of Physics and Chemistry, Chinese Academy of Sciences, 40-1 South Beijing Road, Urumqi 830011, China.}
    \affiliation{Center of Materials Science and Optoelectronics Engineering, University of Chinese Academy of Sciences, Beijing 100049, China.}

    \author{Zhihua Yang}
    \email{zhyang@ms.xjb.ac.cn}
    \affiliation{Research Center for Crystal Materials; CAS Key Laboratory of Functional Materials and Devices for Special Environmental Conditions; Xinjiang Key Laboratory of Functional Crystal Materials; Xinjiang Technical Institute of Physics and Chemistry, Chinese Academy of Sciences, 40-1 South Beijing Road, Urumqi 830011, China.}
    \affiliation{Center of Materials Science and Optoelectronics Engineering, University of Chinese Academy of Sciences, Beijing 100049, China.}

    \author{Shilie Pan}
    \email{slpan@ms.xjb.ac.cn}
    \affiliation{Research Center for Crystal Materials; CAS Key Laboratory of Functional Materials and Devices for Special Environmental Conditions; Xinjiang Key Laboratory of Functional Crystal Materials; Xinjiang Technical Institute of Physics and Chemistry, Chinese Academy of Sciences, 40-1 South Beijing Road, Urumqi 830011, China.}
    \affiliation{Center of Materials Science and Optoelectronics Engineering, University of Chinese Academy of Sciences, Beijing 100049, China.}

    \date{\today}

    \begin{abstract}
        In this work, we assess two widely used scissors-correction schemes for first-principles calculations of second-harmonic generation in representative borate and phosphate ultraviolet nonlinear-optical (UV-NLO) crystals, namely scheme-L [Phys.\ Rev.\ Lett.\ \textbf{63}, 1719 (1989)] and scheme-N [Phys.\ Rev.\ B \textbf{72}, 045223 (2005)].
To enable controlled and numerically robust comparisons, we derive a unified static-limit formulation that avoids spurious divergences and is applicable to both schemes, thereby extending earlier static-limit treatments that were effectively restricted to scheme-L.
Benchmark calculations show that both schemes largely preserve the spectral line shape while mainly rescaling the overall response.
Scheme-N systematically yields 15\%--25\% larger SHG magnitudes than scheme-L, although for some tensor components and experimental datasets scheme-L shows closer agreement with experiment.
We further show that Kleinman symmetry is satisfied in the static limit at the level of the formal theory, whereas apparent violations in practical calculations arise mainly from the numerical approximation used to evaluate generalized derivatives.

    \end{abstract}

    \maketitle
    \newpage

    \section{Introduction}
    \label{sec:introduction}
    Ultraviolet (UV) lasers, including deep-ultraviolet (DUV) sources with wavelengths below 200 nm, are important for a wide range of applications such as photolithography, optical inspection, imaging, and precision metrology.
Among the available approaches to UV light generation, nonlinear frequency conversion in bulk nonlinear-optical (NLO) crystals, particularly second-harmonic generation (SHG), is attractive because it can deliver high efficiency, narrow linewidth, and flexible wavelength access.
Reliable first-principles predictions of SHG are therefore important for elucidating structure--property relations and for screening candidate UV/DUV-NLO crystals.

The theoretical description of SHG in crystals has a long history.
As early as 1963, Butcher and McLean computed SHG coefficients within band theory.\cite{Butcher1963}
Their formulation contained explicit divergences in the static limit.
Subsequent work by Aspnes\cite{Aspnes1972} clarified that, for cubic crystals, the divergent terms cancel when crystal symmetry and time-reversal symmetry are properly enforced, in both the velocity and length gauges.
This analysis also showed that the two-band term, which appears explicitly only in the length gauge, does not represent an independent contribution and can be reorganized into three-band terms using completeness.

A major step forward came from the work of Ghahramani, Sipe, and co-workers in the early 1990s.\cite{Ghahramani1991,Sipe1993,Moss1987,Moss1990}
By exploiting a sum rule and imposing time-reversal symmetry in the velocity gauge, they obtained a general divergence-free formalism for second- and third-order optical responses.
In 1993, Sipe \textit{et al.}\ further refined the theory by systematically separating interband and intraband contributions within a length-gauge framework.\cite{Sipe1993}
They also demonstrated the equivalence between the velocity-gauge formulation of Ghahramani \textit{et al.}\ and the corresponding length-gauge expressions.
Building on this foundation, Aversa and Sipe\cite{Aversa1995} derived general mixing-frequency expressions for $\chi^{(2)}(-\omega_2;\omega_\beta,\omega_\alpha)$ with $\omega_2=\omega_\beta+\omega_\alpha$, free of unphysical divergences.
They further established the equivalence between the length and velocity gauges via a unitary transformation.

In the static limit, Rashkeev \textit{et al.}\ rearranged the length-gauge formalism to make the symmetry properties more transparent and to enforce Kleinman symmetry explicitly.\cite{Kleinman1962,Rashkeev1998}
Subsequent work addressed computational efficiency and numerical stability.
Duan \textit{et al.}\ developed an evaluation technique that substantially reduces the number of $k$ points required for convergence within the Ghahramani--Sipe formalism,\cite{Duan1999} and Lin \textit{et al.}\ later rearranged the terms to remove potentially problematic denominators while adopting the technique of Duan \textit{et al.}\ to further improve convergence.\cite{Lin1999}
A general frequency-dependent length-gauge formulation was subsequently presented by Sipe and Shkrebtii,\cite{Sipe2000} and has since become a widely used framework for practical calculations of the second-order susceptibility in clean, cold semiconductors within the independent-particle approximation.

Another critical aspect of first-principles SHG calculations is the treatment of quasiparticle corrections, which are often incorporated through a scissors correction.
The original scissors-correction approach proposed by Levine and co-workers,\cite{Levine1989,Levine1990,Levine1991} denoted scheme-L, was later revised by Nastos \textit{et al.}\cite{Nastos2005} to yield scheme-N for SHG calculations.
The key distinction between the two schemes lies in how the generalized derivative is treated in the nonlinear-response formulas.
Despite this clarification, both schemes continue to be used in modern electronic-structure codes.
For example, CASTEP\cite{Segall2002,Clark2005} and ABINIT\cite{Gonze2005,Gonze2020} rely on scheme-L for SHG, whereas GPAW,\cite{Mortensen2024} ArchNLO,\cite{Wang2017c,Song2020} and HopTB\cite{HopTB,Wang2017,Wang2019} use scheme-N.
For UV/DUV-NLO crystals, scheme-L has been widely used in practice and often yields static SHG coefficients that compare favorably with experimental benchmarks.
It remains unclear whether scheme-N provides systematically closer agreement with experiment for representative UV/DUV-NLO crystals.

A first goal of this work is to quantify how the choice between scheme-L and scheme-N affects SHG spectra and coefficients in representative UV/DUV-NLO crystals.
We assess both the frequency-dependent response and the zero-frequency limit.
Because the dispersion of the SHG susceptibility well below the band gap is typically weak, the static limit provides a convenient and widely used descriptor of the below-gap response.
To enable controlled comparisons between the two scissors-correction schemes in this regime, we derive a unified and numerically stable static-limit formulation that avoids spurious divergences, accommodates both prescriptions, and enforces Kleinman symmetry explicitly.
We implement this workflow in \texttt{NLOkit} to facilitate systematic and reproducible cross-code SHG diagnostics for UV/DUV-NLO crystals.

A second goal is to resolve an apparent inconsistency regarding Kleinman symmetry in practical sum-over-states (SOS) calculations.
Although the formal static-limit analysis of Rashkeev \textit{et al.}\ yields expressions that satisfy Kleinman symmetry,\cite{Rashkeev1998} several implementations based on the $\omega\!\to\!0$ limit of the frequency-dependent formulas report noticeable violations.\cite{Cheng2019,Cheng2020b,Jia2020}
Here we clarify the origin of these discrepancies and attribute the observed symmetry breaking primarily to the numerical approximation used to evaluate generalized derivatives.

Local-field effects can be important for quantitative SHG predictions and have been studied.\cite{Levine1990,Levine1991,Chen1997,Bertocchi2012}
For semiconductors and insulators, the resulting corrections to SHG are typically at the level of tens of percent (about $-20\%$ to $+30\%$).\cite{Chen1997}
In this work, we neglect local-field effects to isolate the impact of the choice of scissors-correction scheme within the independent-particle SOS framework.
We also note that tight-binding-like approaches based on Wannier functions or nonorthogonal atomic basis sets provide an alternative route to nonlinear-optical properties.\cite{Wang2017,Wang2019,Chen2022b,Wang2024}
A key advantage of these approaches is that the generalized derivative can be evaluated without the large unoccupied-band sums required by SOS methods.\cite{Wang2017}

The outline of the rest of the paper is as follows.
Section~\ref{sec:methods} summarizes the theoretical framework.
Section~\ref{sec:details} provides the computational details.
Results and discussion are presented in Section~\ref{sec:results}, and the main conclusions are summarized in Section~\ref{sec:summary}.

    \section{Methods}
    \label{sec:methods}
    In this section, we assume time-reversal symmetry (TRS) and neglect spin--orbit coupling.

\subsection{SHG formalism and generalized derivatives}
\label{subsec:chi_dyn}

We evaluate the frequency-dependent SHG susceptibility using the independent-particle length-gauge formalism of Rashkeev \textit{et al.}:\cite{Rashkeev1998}
\begin{align}
    \chi^{abc}(-2\omega;\omega,\omega)
    = \chi^{abc}_e(-2\omega;\omega,\omega)
    + \chi^{abc}_i(-2\omega;\omega,\omega),
\end{align}
where $\chi^{abc}_e$ and $\chi^{abc}_i$ denote the purely interband and mixed interband--intraband contributions, respectively.
The complete frequency-dependent expressions, together with the definitions of the relevant quantities and the static-limit derivation, are given in Sec.~S1 of the Supporting Information.
Causality is enforced by replacing $\omega$ with $\omega+i\eta$ in the frequency denominators and taking the limit $\eta\to0^+$.

For distinct, nondegenerate bands, i.e., $n\neq m$ and $\omega_{nm}\neq0$, the interband Berry connection can be written as
\begin{align}
    r_{nm}^a = -i\frac{v_{nm}^a}{\omega_{nm}}
    = \frac{p_{nm}^a}{i m_e\omega_{nm}},
    \label{eq:r_p_relation}
\end{align}
where the second equality assumes the local-Hamiltonian relation $\hat{\mathbf p}=m_e\hat{\mathbf v}$, and $m_e$ is the free-electron mass.
The corresponding velocity difference is
\begin{align}
    \Delta_{mn}^a = v_{mm}^a-v_{nn}^a
    = \frac{p_{mm}^a-p_{nn}^a}{m_e}.
\end{align}
Equation~\eqref{eq:r_p_relation} cannot be applied directly within an exactly degenerate subspace because its denominator vanishes and the individual off-diagonal matrix elements depend on the choice of basis.
For exactly degenerate distinct bands, the wave functions within the degenerate subspace may be chosen such that the relevant off-diagonal matrix elements vanish.\cite{Aversa1995,Rashkeev1998}
In numerical calculations, near-degenerate pairs are treated using the tolerance described below.

The generalized derivative,
\begin{align}
    r^a_{nm;b}=\partial_{k_b}r^a_{nm}
    -i(\xi^b_{nn}-\xi^b_{mm})r^a_{nm},
\end{align}
is evaluated using the sum rule \cite{Sipe1993,Wang2017}
\begin{align}
    r^b_{nm;a}
    &=
    \frac{r_{nm}^{a}\Delta_{mn}^{b}
          +r_{nm}^{b}\Delta_{mn}^{a}}{\omega_{nm}}
    +\frac{i}{\omega_{nm}}\sum_l
      \left(\omega_{lm}r_{nl}^{a}r_{lm}^{b}
           -\omega_{nl}r_{nl}^{b}r_{lm}^{a}\right),
    \label{eq:gen_deriv_sum_rule}
\end{align}
which is exact for local Hamiltonians with complete band sums.
For nonlocal Hamiltonians, an additional term proportional to
$\langle n|\partial_{k_b}\partial_{k_a}\hat H|m\rangle$
may contribute.\cite{Wang2017}

\subsection{Static limit and numerical Kleinman symmetry}
\label{ssec:kleinman_in_static_case}

In the static limit, the mixed contribution can be rearranged into the following manifestly Kleinman-symmetric form:\cite{Rashkeev1998,Chen2022b,Wang2024}
\begin{align}
    \chi^{abc}_{i}
    = \frac{e^{3}}{4\hbar^{2}}P(abc)
    \int \frac{d\mathbf{k}}{(2\pi)^3}
    \sum_{nm}\frac{f_{nm}}{\omega_{mn}^{2}}
    \Im\left\{r^a_{nm}r^b_{mn;c}\right\},
    \label{eq:kleinman_sym}
\end{align}
where $P(abc)$ denotes the full permutation over the Cartesian indices $(a,b,c)$.
The derivation of Eq.~\eqref{eq:kleinman_sym}, together with the corresponding interband expression exhibiting explicit Kleinman symmetry, is provided in Sec.~S1 of the Supporting Information.

Although Kleinman symmetry is exact for the formal static-limit expressions, practical calculations can exhibit residual differences between symmetry-related tensor components.\cite{Cheng2019,Cheng2020b,Jia2020}
The main numerical sources of these residual differences are
(i) finite $\mathbf{k}$-point integration,
(ii) truncation of the intermediate-state sum in Eq.~\eqref{eq:gen_deriv_sum_rule},
(iii) near-degenerate denominators introduced when Eq.~\eqref{eq:r_p_relation} is used, and
(iv) nonlocal-potential contributions that are not included in the local-Hamiltonian sum rule.

To treat near-degenerate band-energy denominators, we introduce an energy tolerance $\epsilon$.
When constructing $r^a_{nm}$ from Eq.~\eqref{eq:r_p_relation}, we set $r^a_{nm}=0$ whenever $|\omega_{nm}|<\epsilon$.
In the three-band part of Eq.~\eqref{eq:gen_deriv_sum_rule}, we omit an intermediate-state term whenever $|\omega_{nl}|<\epsilon$ or $|\omega_{lm}|<\epsilon$.
Equivalently, when the intermediate band $l$ is exactly degenerate with $n$ or $m$, the basis within the degenerate subspace can be chosen such that the relevant factors $p^a_{nl}$ or $p^a_{lm}$, and therefore the corresponding $r^a_{nl}$ or $r^a_{lm}$, vanish.\cite{Aversa1995,Rashkeev1998}
This prescription removes numerically singular contributions associated with small energy denominators and is consistent with setting the corresponding position-matrix element to zero in the exactly degenerate limit.
The tolerance is also important for obtaining smooth and well-converged SHG spectra.\cite{Song2020}

To diagnose these numerical effects, we decompose the unsymmetrized mixed contribution into a two-band part, $\chi^{abc}_{i,\mathrm{two}}$, obtained by omitting the intermediate-state sum over $l$ in Eq.~\eqref{eq:gen_deriv_sum_rule}, and a residual three-band part,
$\chi^{abc}_{i,\mathrm{three}}=\chi^{abc}_i-\chi^{abc}_{i,\mathrm{two}}$.
This diagnostic decomposition does not enforce Kleinman symmetry term by term and is used in Sec.~\ref{sec:results} to analyze conduction-band convergence.

\subsection{Scissors-correction schemes}

For the frequency-dependent response, both scissors-correction protocols rigidly shift the conduction-band energies.
In scheme N,\cite{Nastos2005} the generalized derivative is kept unchanged, so the shift enters only the response denominators.
In scheme L,\cite{Levine1989,Levine1990,Levine1991} the shift also modifies the transition energies used in Eq.~\eqref{eq:gen_deriv_sum_rule}.

For the static response, we use an explicitly Kleinman-symmetric formulation that accommodates both schemes.
We first decompose the susceptibility as
\begin{align}
    \chi^{abc}
    &= \chi^{abc}_{i}+\chi^{abc}_{e}
     = \chi^{abc}_{\mathrm{II}}+\chi^{abc}_{i,\mathrm{III}}
       +\chi^{abc}_{e}
     = \chi^{abc}_{\mathrm{II}}+\chi^{abc}_{\mathrm{III}},
\end{align}
where
\begin{align}
    \chi^{abc}_{\mathrm{III}}
    =\chi^{abc}_{i,\mathrm{III}}+\chi^{abc}_{e}.
\end{align}
The intermediate derivation is given in Sec.~S1 of the Supporting Information.
The two-band term is
\begin{align}
\chi^{abc}_\mathrm{II}
&= \frac{i e^{3}}{2\hbar^{2}}
   \int \frac{d\mathbf{k}}{(2\pi)^3}\sum_{nm}
   \frac{f_{nm}}{\omega_{mn}^{3}}
   P(abc)r^{a}_{nm}r^{b}_{mn}\Delta^{c}_{nm},
   \label{eq:chi_II}
\end{align}
or, after restricting the sums to valence ($V$) and conduction ($C$) bands,
\begin{align}
\chi^{abc}_\mathrm{II}
&= \frac{e^{3}}{\hbar^{2}}
   \int \frac{d\mathbf{k}}{(2\pi)^3}
   \sum_{n\in V}\sum_{m\in C}
   \frac{f_{nm}}{\omega_{mn}^{5}}P(abc)
   \Im\left[p^{a}_{nm}p^{b}_{mn}
   \left(p^{c}_{nn}-p^{c}_{mm}\right)\right].
   \label{eq:two-band-contr}
\end{align}
Under TRS, this term vanishes in the static limit,\cite{Zhang2015} so the nonzero static response is governed by $\chi^{abc}_{\mathrm{III}}$.
Nevertheless, $\chi^{abc}_{\mathrm{II}}$ is retained here because it appears naturally in the formal decomposition and provides a useful check of the static-limit reduction.

We denote the scissors shift in frequency units by $\Delta$ and define the scissors-corrected transition frequency as
\begin{align}
    S_{nm}=\omega_{nm}+f_{nm}\Delta,
\end{align}
with the same sign convention as that used for $\omega_{nm}$ and $f_{nm}$.
In scheme N, the three-band contribution is
\begin{align}
\chi^{abc}_\mathrm{III,N}
&= \frac{e^{3}}{2\hbar^{2}}
   \int \frac{d\mathbf{k}}{(2\pi)^3}
   \sum_{n\in V}\sum_{m\in C}\sum_{l\in V}
   P(abc)\Im\left\{p^{a}_{nm}p^{b}_{ml}p^{c}_{ln}\right\}
   \frac{1}{S_{mn}^2\omega_{nm}\omega_{lm}}
   \left(\frac{1}{S_{lm}}+\frac{2}{\omega_{nm}}\right)
   \nonumber\\
&\quad
+\frac{e^{3}}{2\hbar^{2}}
   \int \frac{d\mathbf{k}}{(2\pi)^3}
   \sum_{n\in V}\sum_{m\in C}\sum_{l\in C}
   P(abc)\Im\left\{p^{a}_{nm}p^{b}_{ml}p^{c}_{ln}\right\}
   \frac{1}{S_{mn}^2\omega_{mn}\omega_{ln}}
   \left(\frac{1}{S_{ln}}+\frac{2}{\omega_{mn}}\right).
   \label{eq:highlight}
\end{align}
In Eq.~\eqref{eq:highlight}, the same degenerate-state convention applies to momentum factors connecting states within the valence or conduction subspace.
Specifically, if $l$ is exactly degenerate with $m$ or $n$, the basis can be chosen such that the corresponding factor $p^{b}_{ml}$ or $p^{c}_{ln}$ vanishes, and the term therefore makes no contribution.

In scheme L, the corresponding three-band contribution is
\begin{align}
\chi^{abc}_\mathrm{III,L}
&= \frac{e^{3}}{2\hbar^{2}}
   \int \frac{d\mathbf{k}}{(2\pi)^3}
   \sum_{n\in V}\sum_{m\in C}\sum_{l\in V}
   P(abc)\Im\left\{\tilde p^{a}_{nm}\tilde p^{b}_{ml}
   \tilde p^{c}_{ln}\right\}
   \left(\frac{1}{S_{nm}^3S_{lm}^2}
        +\frac{2}{S_{nm}^4S_{lm}}\right)
   \nonumber\\
&\quad
+\frac{e^{3}}{2\hbar^{2}}
   \int \frac{d\mathbf{k}}{(2\pi)^3}
   \sum_{n\in V}\sum_{m\in C}\sum_{l\in C}
   P(abc)\Im\left\{\tilde p^{a}_{nm}\tilde p^{b}_{ml}
   \tilde p^{c}_{ln}\right\}
   \left(\frac{1}{S_{mn}^3S_{ln}^2}
        +\frac{2}{S_{mn}^4S_{ln}}\right),
   \label{eq:vh_ve}
\end{align}
where
\begin{align}
    \tilde p^a_{nm}
    = \left[1+\frac{\Delta}{\omega_{nm}}
      \left(\delta_{nC}-\delta_{mC}\right)\right]p^a_{nm}.
\end{align}
Here, $\delta_{nC}=1$ if $n$ is a conduction band and zero otherwise.
The same convention is used in Eq.~\eqref{eq:vh_ve}: $\tilde p^{b}_{ml}$ or $\tilde p^{c}_{ln}$ is set to zero when $l$ is exactly degenerate with $m$ or $n$ within the same valence or conduction subspace.
The first and second terms in Eq.~\eqref{eq:vh_ve} correspond to the VH and VE terms, respectively, of Ref.~\citenum{Lin1999}.

When $\Delta=0$, one has $S_{nm}=\omega_{nm}$ and $\tilde p^a_{nm}=p^a_{nm}$, so Eqs.~\eqref{eq:highlight} and \eqref{eq:vh_ve} reduce to the same uncorrected static-limit expression.
The distinction between schemes N and L therefore arises only for a finite scissors shift.

    \section{Computational Details}
    \label{sec:details}
    In this work, to assess how scissors-correction schemes affect SHG predictions, we studied five representative borate UV/DUV-NLO crystals: $\beta$-\ce{BaB2O4} (BBO),\cite{Chen1985} \ce{LiB3O5} (LBO),\cite{Chen1989a} \ce{CsB3O5} (CBO),\cite{Wu1993} \ce{CsLiB6O10} (CLBO),\cite{Mori1995} and \ce{KBe2BO3F2} (KBBF),\cite{Mei1995,Wu1996} together with the phosphate \ce{LiCs2PO4} (LCPO).\cite{Li2016c,Shen2016}
To test the hypothesis that apparent violations of Kleinman symmetry are primarily numerical in origin, we additionally examined \ce{KH2PO4} in its paraelectric phase [P-KDP, space group $I\overline{4}2d$, point group $\overline{4}2m$] and ferroelectric phase [F-KDP, space group $Fdd2$, point group $mm2$].\cite{Jia2020}
We further included \ce{BaNa2[PO3(OH)]2} (BNPO, space group $Fdd2$, point group $mm2$),\cite{Yang2023} which was not considered in Ref.~\citenum{Jia2020}.
Because BNPO has the same lower-symmetry point group as F-KDP but a different composition and structure, it provides an independent test of whether the enhanced numerical Kleinman-symmetry breaking is associated with lower crystallographic symmetry.

For BBO, LBO, CBO, CLBO, KBBF, and LCPO, first-principles calculations were performed with CASTEP.\cite{Segall2002,Clark2005}
For P-KDP, F-KDP, and BNPO, we carried out cross-code comparisons using CASTEP, the Vienna \textit{ab initio} Simulation Package (VASP)\cite{Kresse1996a}, and the projector-augmented-wave code GPAW.\cite{Mortensen2024}
All calculations were based on density-functional theory (DFT) within the generalized-gradient approximation using the Perdew--Burke--Ernzerhof (PBE) exchange--correlation functional.\cite{Perdew1996}
Projector-augmented-wave (PAW) datasets were used in VASP~5.4.4 and GPAW~25.7.0, whereas norm-conserving pseudopotentials\cite{Lin1993,Lee1995} were used in CASTEP.
The cross-code calculations used comparable valence spaces for H, O, P, K, and Ba.
In CASTEP and VASP, H $1s^1$, O $2s^2 2p^4$, P $3s^2 3p^3$, K $3s^2 3p^6 4s^1$, Na $2s^2 2p^6 3s^1$, and Ba $5s^2 5p^6 6s^2$ were treated as valence states.
In GPAW, the default PBE PAW setups were used, for which H $1s^1$, O $2s^2 2p^4$, P $3s^2 3p^3$, K $3s^2 3p^6 4s^1$, Na $2p^6 3s^1$, and Ba $5s^2 5p^6 6s^2$ are included as valence states.

The structures of BBO, LBO, CBO, CLBO, KBBF, and LCPO were taken from the PBE-optimized geometries reported in Refs.~\cite{Cheng2018a,Cheng2020b}.
The PBE-optimized geometries of P-KDP and F-KDP were taken from Ref.~\cite{Jia2020}.
For BNPO, the experimental structure reported in Ref.~\cite{Yang2023} was used as the starting point and was further relaxed within PBE using VASP, including the van der Waals correction suggested in Ref.~\cite{Yang2023}.
For each of P-KDP, F-KDP, and BNPO, identical lattice vectors and atomic coordinates were used in the CASTEP, GPAW, and VASP calculations; no package-specific structural relaxation was performed.

In CASTEP, scalar-relativistic effects were included via the Koelling--Harmon treatment.\cite{Koelling1977}
Self-consistent calculations were converged to $10^{-6}$~eV per atom, except for the F-KDP CASTEP cross-check, where a threshold of $10^{-10}$~eV per atom was used.
The plane-wave kinetic-energy cutoff was set to 750~eV for BBO, LBO, CBO, CLBO, LCPO, P-KDP, and BNPO, to 850~eV for KBBF, and to 1000~eV for the F-KDP CASTEP cross-check.
Brillouin-zone integrations used Monkhorst--Pack meshes of $6\times6\times6$ (28), $6\times7\times9$ (60), $11\times8\times7$ (96), $9\times9\times9$ (75), $13\times13\times13$ (231), $9\times9\times6$ (75), $13\times13\times13$ (196), $6\times6\times6$ (108), and $8\times8\times8$ (95) for BBO, LBO, CBO, CLBO, KBBF, LCPO, P-KDP, F-KDP, and BNPO, respectively, where the numbers in parentheses denote irreducible $k$ points.

Using the converged Kohn--Sham wave functions, we carried out optical calculations on the same $k$-point meshes but included additional unoccupied states to compute momentum-matrix elements, as required by the sum-over-states and sum-rule frameworks.
The numbers of occupied (total) bands used for BBO, LBO, CBO, CLBO, KBBF, LCPO, P-KDP, F-KDP, and BNPO were 120 (569), 80 (492), 96 (724), 88 (453), 24 (162), 48 (269), 40 (904), 40 (1374), and 88 (629), respectively.
For P-KDP, F-KDP, and BNPO, we further varied the $k$-point mesh and the number of unoccupied bands to quantify numerical convergence; these results are discussed in Sec.~\ref{sec:results}.

In the VASP and GPAW cross-checks, Brillouin-zone integrations used Monkhorst--Pack meshes including the $\Gamma$ point.
In GPAW, calculations were performed in plane-wave mode with a kinetic-energy cutoff of 700~eV and $k$-point meshes of $8\times8\times8$ for F-KDP, $13\times13\times13$ for P-KDP, and $8\times8\times8$ for BNPO.
Van der Waals corrections were not included in the GPAW cross-check.
Default numerical precision settings were used.
Specifically, self-consistency was enforced through simultaneous convergence of the total energy, electron density, and Kohn--Sham eigenstates.
The maximum change in total energy over the last three electronic iterations was required to be below $5\times10^{-4}$~eV per valence electron.
The maximum integral of the absolute change in the electron density was required to be below $1\times10^{-4}$ electrons per valence electron.
The maximum integral of the absolute change in the eigenstates was required to be below $4\times10^{-8}$~eV$^{2}$ per valence electron.
For the momentum-matrix calculations in GPAW, 1010 bands were included for P-KDP and F-KDP, and 1084 bands were included for BNPO.

In VASP, self-consistent calculations were performed with a plane-wave cutoff of 700~eV and $k$-point meshes of $8\times8\times8$ for F-KDP and $13\times13\times13$ for P-KDP.
For BNPO, a cutoff of 850~eV and an $8\times8\times8$ $k$-point mesh were employed, together with a van der Waals correction.
The \texttt{Accurate} precision setting was used throughout.
For the momentum-matrix calculations in VASP, 1152 bands were included in all cases.
For P-KDP and F-KDP, electronic self-consistency in the source charge-density calculations was achieved by requiring the change in total energy between successive electronic iterations to be below $10^{-4}$~eV; the corresponding BNPO threshold was $10^{-8}$~eV.
Tightening this threshold to $10^{-7}$~eV for F-KDP produced negligible changes in the calculated SHG tensor elements, confirming convergence with respect to the electronic stopping criterion.

The only identified difference in the valence configurations is Na in BNPO:
CASTEP and VASP include the Na $2s2p$ semicore shell, whereas the standard distributed GPAW Na semicore setup includes Na $2p$ but keeps Na $2s$ frozen.
Thus, the BNPO cross-code comparison should be regarded as a controlled comparison of numerical implementations with a disclosed valence-space difference, rather than as a strictly like-for-like pseudopotential test.

SHG coefficients, including frequency-dependent spectra and static-limit values, were evaluated using our in-house Python package \texttt{NLOkit}.
Currently, \texttt{NLOkit} supports outputs from CASTEP, GPAW, and VASP.
For CASTEP, \texttt{NLOkit} was used to read the momentum-matrix elements stored in the \texttt{.ome\_bin} file produced by the optical calculations.
In GPAW, we used a modified implementation that evaluated momentum-matrix elements while explicitly accounting for crystal symmetries, rather than relying only on time-reversal symmetry as in the official release.
In VASP, we modified the \texttt{OPTIC} module to export the full set of momentum-matrix elements.
For the frequency-dependent spectra, resonant frequency denominators were evaluated at $\omega+i\eta$ using a damping parameter of $\eta=0.035$~eV.
This parameter broadens the resonances and sets their spectral resolution.
Unless otherwise stated, the energy-denominator tolerance was set to $\epsilon=10^{-4}$~Ha.
Terms involving band-energy differences smaller than $\epsilon$ were omitted, including when constructing the interband position matrix elements and evaluating the generalized derivatives.
Although both parameters regularize potentially singular numerical expressions, $\eta$ acts on the optical-frequency denominators, whereas $\epsilon$ acts on band-energy denominators.
Additional static cross-code calculations using $\epsilon=10^{-7}$~Ha are identified below and in the supplementary material.

    \section{Results}
    \label{sec:results}
    Throughout this section, we report the SHG response in terms of the Cartesian second-order susceptibility $\chi^{abc}(-2\omega;\omega,\omega)$ and the contracted Voigt coefficients $d_{ij}$.
Here, $i=1,2,3$ correspond to $a=x,y,z$, and $j=1,\ldots,6$ correspond to the symmetrized index pairs $(bc)=(xx,yy,zz,yz,zx,xy)$.
Within this convention, $d_{ij}=\tfrac{1}{2}\chi^{abc}$, with $a\leftrightarrow i$ and the symmetrized pair $(bc)\leftrightarrow j$.

\subsection{Band-gap comparison and scissors corrections}

Table~\ref{tbl:band_gap} summarizes the PBE band gaps of the nine crystals considered here (column ``Calc.'') and the corresponding experimental gaps (column ``Exp.'') used to define the scissors shifts.
For BBO, LBO, CBO, CLBO, KBBF, and LCPO, the present PBE gaps agree well with those reported in Refs.~\citenum{Cheng2018a,Cheng2020b} (column ``Ref.''), with deviations typically within $0.15$~eV.
Semilocal density-functional approximations such as PBE usually underestimate the band gaps of semiconductors and insulators.
Accordingly, we apply a rigid scissors correction, $\Delta = E_g^{\mathrm{exp}} - E_g^{\mathrm{PBE}}$, to the conduction bands when evaluating the SHG response.
For the present set of compounds, $\Delta$ ranges from $1.5$ to $2.7$~eV.
For P-KDP, F-KDP, and BNPO, the band gaps obtained with CASTEP, GPAW, and VASP are mutually consistent, providing a controlled baseline for cross-code comparisons.
In addition, the VASP band gaps for P-KDP and F-KDP agree well with those reported in Ref.~\citenum{Jia2020}.

\begin{table}[htbp]
  \caption{
    Computed and experimental band gaps of BBO, LBO, CBO, CLBO, KBBF, LCPO, P-KDP, F-KDP, and BNPO.
    Shown are the GGA-PBE band gaps obtained in this work (column ``Calc.'') and those reported in Refs.~\citenum{Cheng2018a,Cheng2020b} (column ``Ref.'').
    The experimental band gaps (column ``Exp.'') collected from the literature are also listed, together with the values adopted in Refs.~\citenum{Cheng2018a,Cheng2020b} for the scissors corrections.
    All band gaps are given in eV.
  }
  \centering
  \begin{tabular}{lccc}
    \toprule
    Material & Calc. & Ref. & Exp. \\
    \midrule
    BBO  & 4.710 & 4.800~\cite{Cheng2020b} & 6.716~\cite{Cheng2020b} (185~nm\cite{Chen2009a}, 190~nm\cite{Chen1985}, 193~nm\cite{French1991}, 195~nm\cite{Chen2012-ch3}) \\
    LBO  & 6.373 & 6.382~\cite{Cheng2020b} & 8.283~\cite{Cheng2020b} (155~nm\cite{Chen2012-ch3}, 150~nm\cite{Chen2009a}) \\
    CBO  & 5.368 & 5.341~\cite{Cheng2020b} & 7.439~\cite{Cheng2020b} (167~nm\cite{Kagebayashi1999}) \\
    CLBO & 5.041 & 5.093~\cite{Cheng2020b} & 6.902~\cite{Cheng2020b} (180~nm\cite{Chen2012-ch3}) \\
    KBBF & 6.093 & 6.070~\cite{Cheng2020b} & 8.452~\cite{Cheng2020b} (147~nm\cite{Chen2012-ch3}, 155~nm\cite{Wu1996}) \\
    LCPO & 4.296 & 4.43~\cite{Cheng2018a}  & 7.02~\cite{Cheng2020b}  (174~nm\cite{Li2016c}, 190~nm\cite{Shen2016}) \\
    P-KDP & 5.408,\footnotemark[1] 5.632,\footnotemark[2] 5.607,\footnotemark[3] & 5.61\cite{Jia2020}\footnotemark[3] & 7.12\cite{Dmitriev1999} \\
    F-KDP & 5.252,\footnotemark[1] 5.479,\footnotemark[2] 5.451,\footnotemark[3] & 5.45\cite{Jia2020}\footnotemark[3] & 8.0\cite{Ogorodnikov2001} \\
    BNPO  & 4.852,\footnotemark[1] 5.066,\footnotemark[2] 5.025,\footnotemark[3] & $--$ & 6.525 (190~nm)~\cite{Yang2023} \\
    \bottomrule
  \end{tabular}
  \footnotetext[1]{Computed with CASTEP.}
  \footnotetext[2]{Computed with GPAW.}
  \footnotetext[3]{Computed with VASP.}
  \label{tbl:band_gap}
\end{table}

\subsection{Linear dielectric response at 1064~nm}

As an independent check of the empirical scissors shifts, we also computed the linear electronic dielectric response at 1064~nm both without a scissors correction and with the gap-derived scissors shift.
Each dielectric tensor was converted to its principal refractive indices and compared with available experimental room-temperature or near-room-temperature refractive-index data at 1064~nm.
The linear-response calculations were performed with \texttt{NLOkit}.
For this first-order response, scheme-L and scheme-N are identical because both reduce to the same rigid shift of the conduction-band energies.\cite{Nastos2005}
The comparison includes BBO, LBO, CBO, CLBO, and KBBF, with experimental refractive-index data taken from Refs.~\citenum{Eimerl1987,Chen1989a,Hanson1991LBO,Zhang2013CBO,Sasaki2003CLBO,Li2016KBBF}.
The corresponding principal-index data for LCPO were not found; therefore, LCPO is included only in the calculated tensor table in the supplementary material.

Table~\ref{tbl:dielectric_metrics} summarizes the mean absolute error (MAE) and root-mean-square error (RMSE) of the calculated refractive indices.
The uncorrected PBE calculation overestimates the refractive indices, giving MAE($n$) = 0.084 and RMSE($n$) = 0.085.
Applying the same gap-derived scissors shifts used in the SHG calculations reduces these errors to MAE($n$) = 0.015 and RMSE($n$) = 0.016.
Thus, for the crystals where experimental dielectric data are available, the gap-derived shifts improve the independent linear-optical comparison rather than merely tuning the nonlinear response.
Because this comparison is made at a single transparent wavelength and does not include local-field or excitonic effects, we do not refit the scissors shifts to the dielectric constants.
A dielectric-fitted shift would instead provide a useful sensitivity test, rather than replacing the gap-derived shifts used below.
Full tensor data and source notes are provided in Table~S1 of the supplementary material.

\InputIfFileExists{dielectric_metrics.itex}{}{\message{LaTeX Warning: File `\filename@base.\ifx\filename@ext\relax tex\else\filename@ext\fi' not found on input line \the\inputlineno}

\subsection{Frequency-dependent SHG response}

Figure~\ref{fig:bbo} shows the frequency-dependent SHG spectra $|\chi^{abc}(-2\omega;\omega,\omega)|$ of BBO computed without a scissors correction and with the two scissors-correction schemes, scheme-N and scheme-L.
The corresponding spectra for LBO, CBO, CLBO, KBBF, and LCPO are provided in Figs.~S1--S5 of the supplementary material.
The imaginary parts of representative tensor components for all six crystals are shown in Fig.~S6 of the supplementary material.
For all tensor components and all test crystals considered here, the two scissors-correction schemes preserve the overall spectral line shape, while scheme-N gives larger amplitudes than scheme-L.
The same trend appears in the static limit and persists throughout the finite-frequency range.
This behavior indicates that, for the present systems, the two schemes primarily rescale the magnitude of the nonlinear response, with only minor changes in its frequency dependence, consistent with Ref.~\citenum{Nastos2005}.

For BBO, the dominant contribution is $\chi^{yyy}$, equivalently $d_{22}$, whereas $\chi^{xxz}$, $\chi^{zxx}$, and $\chi^{zzz}$ are much smaller.
The components $\chi^{xxz}$ and $\chi^{zxx}$ are nearly identical in the static limit but become distinct at finite frequencies.
The separation becomes pronounced above $\sim 4$~eV, indicating the progressive breakdown of full Kleinman symmetry as resonant interband transitions are approached.
Similar behavior is observed for the corresponding component pairs in the other compounds, as shown in Figs.~S1--S5.

\begin{figure}[h]
    \centering
    \includegraphics[scale=0.83]{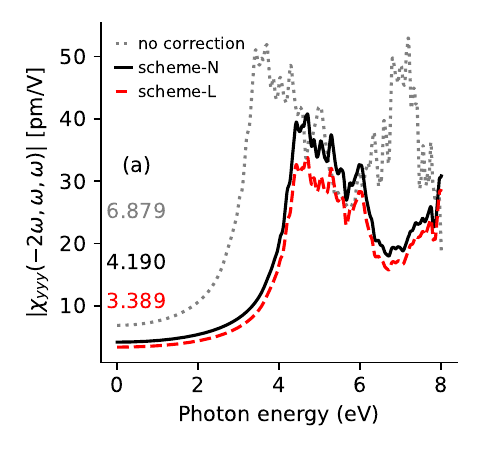}
    \includegraphics[scale=0.83]{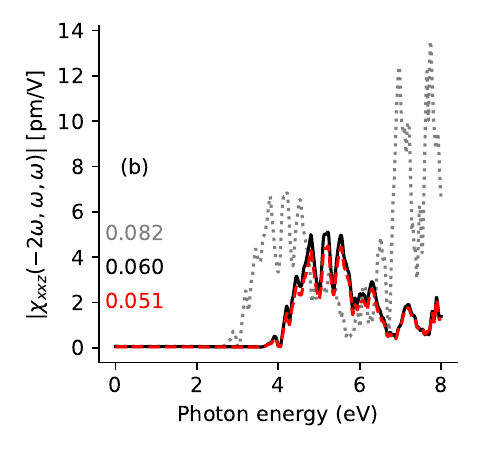}\\
    \includegraphics[scale=0.83]{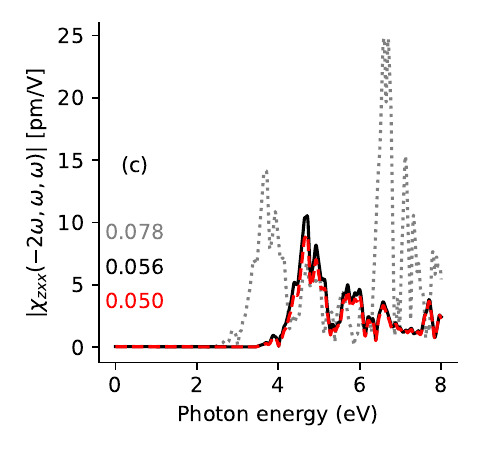}
    \includegraphics[scale=0.83]{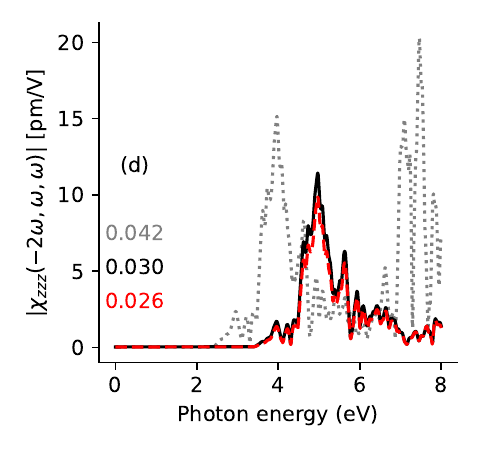}
    \caption{
    Frequency-dependent SHG spectra $|\chi^{abc}(-2\omega;\omega,\omega)|$ of BBO as functions of the photon energy $\hbar\omega$ for (a) $\chi^{yyy}$, (b) $\chi^{xxz}$, (c) $\chi^{zxx}$, and (d) $\chi^{zzz}$.
    Results are shown without a scissors correction (gray dotted lines) and with the two scissors-correction schemes, scheme-N (black solid lines) and scheme-L (red dashed lines).
    The values printed near $\hbar\omega=0$ are the corresponding static-limit coefficients obtained with each scheme.
    A damping parameter of $\eta=0.035$~eV was used in the resonant frequency denominators.
    }
    \label{fig:bbo}
\end{figure}

\subsection{Static SHG coefficients and comparison with experiment}

Table~\ref{tbl:static_shg} lists the SHG coefficients $d_{ij}$, in pm/V, for BBO, LBO, CBO, CLBO, KBBF, and LCPO in the static limit ($\omega=0$) and at $\omega=1.165$~eV, corresponding to 1064~nm.
In the static limit, we report both the unsymmetrized tensor (``unsym'') and the Kleinman-symmetrized tensor (``sym'').
The two sets of values differ only slightly, typically at the sub-percent to few-percent level, indicating small numerical deviations from Kleinman symmetry and good convergence in the static limit.
At $\omega=1.165$~eV, we report only the unsymmetrized values.
Because this photon energy lies well below the absorption onset of these wide-gap crystals, the response is predominantly real; therefore, we report $\mathrm{Re}\,d_{ij}$.

For the static SHG coefficients obtained with scheme-L, the calculated magnitudes agree with the previously reported scheme-L unsymmetrized results [column ``Ref.~\citenum{Cheng2020b}''].
Residual sign differences may arise from implementation-dependent sign conventions in the optical-response formalism, for example from the choice of $e$ or $-e$ in the electron-field coupling.
Such differences do not affect magnitude-based comparisons, although sign-sensitive comparisons require a consistent convention.
Replacing scheme-L with scheme-N systematically increases the magnitudes of $d_{ij}$ for all compounds and tensor components listed in Table~\ref{tbl:static_shg}.
To quantify this protocol dependence, we define
\begin{equation}
p_{ij}(\mathrm{X})=\frac{d_{ij}^{(\mathrm{N})}-d_{ij}^{(\mathrm{L})}}{d_{ij}^{(\mathrm{N})}}\times 100\%,
\end{equation}
where $d_{ij}^{(\mathrm{L})}$ and $d_{ij}^{(\mathrm{N})}$ are obtained with scheme-L and scheme-N, respectively, and $\mathrm{X}$ denotes the compound.
When sign conventions differ, the same expression is applied to $|d_{ij}|$.
For the largest-magnitude coefficient in each crystal, we obtain $p_{22}(\mathrm{BBO})=19.12\%$, $p_{32}(\mathrm{LBO})=16.34\%$, $p_{14}(\mathrm{CBO})=17.44\%$, $p_{36}(\mathrm{CLBO})=17.29\%$, $p_{11}(\mathrm{KBBF})=18.50\%$, and $p_{15}(\mathrm{LCPO})=25.55\%$.
These values are consistent with the enhancement reported in Ref.~\citenum{Nastos2005}.

Where experimental data at 1064~nm are available (Exp.\ column), the reported values exhibit a noticeable spread, particularly for the borates.
Within this range, scheme-L generally provides a lower estimate and often lies near the lower bound of the experimental values, as in the case of $d_{11}$ for KBBF.
By contrast, scheme-N shifts the calculated values upward and, for some components in BBO and LBO, yields better agreement with the larger experimental values.
This trend is shown more clearly in Fig.~\ref{fig:shg_vs_exp}, which compares the calculated and experimental SHG coefficients at 1064~nm for the largest tensor component of each crystal using the experimental datasets of Refs.~\citenum{Dmitriev1999,Chen2012-ch3}.
The MAE and RMSE values reported in the figure quantify the deviations from experiment.
For the dataset of Ref.~\citenum{Dmitriev1999}, scheme-N yields smaller MAE and RMSE values, whereas for the dataset of Ref.~\citenum{Chen2012-ch3}, scheme-L yields slightly smaller errors.
Overall, Table~\ref{tbl:static_shg} and Fig.~\ref{fig:shg_vs_exp} show that the choice of scissors-correction scheme produces a systematic change of about 15\%--25\% in the SHG magnitude while preserving the relative ordering of tensor components within each compound.
At the same time, the question of which scheme is closer to experiment remains affected by the non-negligible spread in the available measurements.

\InputIfFileExists{shg_tbl.itex}{}{\message{LaTeX Warning: File `\filename@base.\ifx\filename@ext\relax tex\else\filename@ext\fi' not found on input line \the\inputlineno}

\begin{figure}[h]
    \centering
    \includegraphics[scale=1.0]{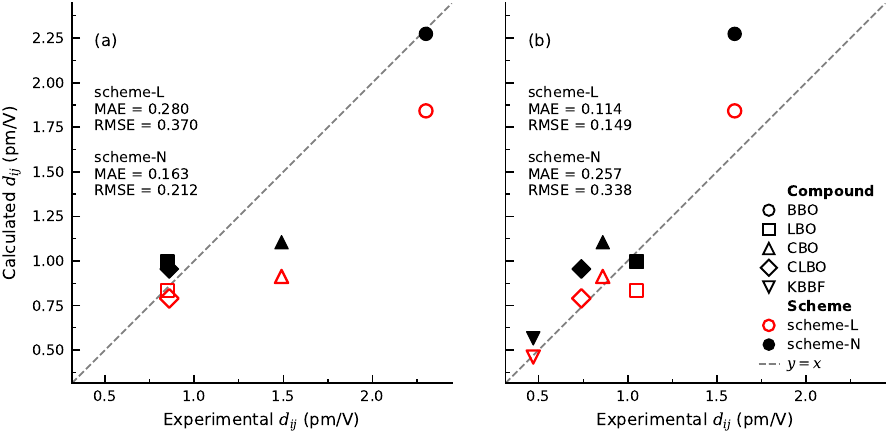}
    \caption{
    Calculated versus experimental SHG coefficients at 1064~nm for the largest tensor component of each crystal.
    Panels (a) and (b) use the experimental data from Refs.~\citenum{Dmitriev1999} and \citenum{Chen2012-ch3}, respectively.
    The dashed line denotes perfect agreement.
    Marker shape identifies the compound, while open red and filled black symbols denote the scheme-L and scheme-N results, respectively.
    The insets list the mean absolute error (MAE) and root-mean-square error (RMSE) of each scheme with respect to the corresponding experimental dataset.
    Absolute values are plotted.
    }
    \label{fig:shg_vs_exp}
\end{figure}

\subsection{Cross-code assessment of numerical Kleinman symmetry}

We next use P-KDP, F-KDP, and BNPO to examine the apparent breaking of Kleinman symmetry in practical static-limit calculations.
P-KDP and F-KDP were studied in Ref.~\citenum{Jia2020}, which reported unequal values for tensor components that should be identical under Kleinman symmetry.
BNPO was not included in that work.
We include BNPO because it has the same lower-symmetry point group $mm2$ as F-KDP, whereas P-KDP belongs to the higher-symmetry point group $\overline{4}2m$.
The comparison between F-KDP and BNPO therefore provides an independent test of whether lower crystallographic symmetry is associated with greater numerical sensitivity.
The purpose of the CASTEP, GPAW, and VASP comparison is not to define three independent primary datasets, but to determine whether the residual symmetry mismatch is robust or depends on the electronic-structure and optical-matrix implementation.

Table~\ref{tbl:shg} compares the static-limit coefficients obtained with CASTEP, GPAW, and VASP with the representative VASP results of Ref.~\citenum{Jia2020}.
Unless otherwise stated, we use $\epsilon=10^{-4}$~Ha.
Results obtained with $\epsilon=10^{-7}$~Ha are given in Table~S2 of the supplementary material.
We use scheme-L to match the setup of Ref.~\citenum{Jia2020}.
To quantify the deviation between a pair $(d_{ij},d_{kl})$, we adopt the relative mismatch used in Ref.~\citenum{Jia2020}:
\begin{align}
    \Delta(d_{ij},d_{kl})=\frac{2(d_{ij}-d_{kl})}{d_{ij}+d_{kl}}\times 100\%.
\end{align}
Here, $d_{ij}-d_{kl}$ is the difference between the two symmetry-related coefficients, and $(d_{ij}+d_{kl})/2$ is their average.
The latter corresponds to the value obtained when Kleinman symmetry is enforced by symmetrization.
Although symmetrized values can be reported explicitly, as in Table~\ref{tbl:static_shg} for the other compounds, we retain the definition of Ref.~\citenum{Jia2020} for direct comparison.

For P-KDP, the CASTEP and GPAW values are close, and their magnitudes are slightly smaller than the corresponding VASP values.
Relative to the VASP reference values reported in Ref.~\citenum{Jia2020} ($d_{14}=0.505$ and $d_{36}=0.490$~pm/V), our VASP results ($d_{14}=-0.581$ and $d_{36}=-0.579$~pm/V) are larger in magnitude.
This difference mainly reflects the larger conduction-band energy window adopted here, which increases the number of bands entering the SHG summations.
In particular, the conduction-band cutoff used in this work is 178.31~eV, compared with 25~eV in Ref.~\citenum{Jia2020}.
Despite these differences in magnitude, $d_{14}$ and $d_{36}$ show excellent internal consistency across all three implementations.
The relative mismatch is $\Delta(d_{14},d_{36})=1.14\%$ for CASTEP, $1.31\%$ for GPAW, and $0.34\%$ for VASP, comparable to the reference value of $3.02\%$.\cite{Jia2020}

For F-KDP, CASTEP and GPAW again give similar values, whereas the VASP results show stronger component dependence.
For example, $d_{15}$ from CASTEP and GPAW, 0.325 and 0.340~pm/V, respectively, is smaller than the VASP value of 0.437~pm/V.
By contrast, $d_{31}$ from CASTEP and GPAW, 0.350 and 0.354~pm/V, respectively, is slightly larger than the VASP value of 0.340~pm/V.
Our VASP values are close to those reported in Ref.~\citenum{Jia2020}, except for $d_{31}$, which is also likely sensitive to the conduction-band cutoff.
Here, the conduction-band cutoff is 176.34~eV, whereas Ref.~\citenum{Jia2020} used 25~eV.
The symmetry-related pairs show increased sensitivity to the implementation.
CASTEP and GPAW yield modest mismatches, with $\Delta(d_{15},d_{31})=7.41\%$ and $4.03\%$, and $\Delta(d_{24},d_{32})=8.28\%$ and $0.81\%$, respectively.
VASP yields substantially larger mismatches, with $\Delta(d_{15},d_{31})=24.97\%$ and $\Delta(d_{24},d_{32})=35.73\%$.
These values remain smaller than those reported in Ref.~\citenum{Jia2020}, where $\Delta(d_{15},d_{31})=45.07\%$ and $\Delta(d_{24},d_{32})=55.32\%$.
The main difference between the two VASP datasets is the number of conduction bands included in the SHG summations.

For BNPO, the three codes give similar $d_{15}$ values: 0.428, 0.406, and 0.419~pm/V for CASTEP, GPAW, and VASP, respectively.
For $d_{31}$, the VASP value of 0.471~pm/V is larger than the CASTEP value of 0.425~pm/V, whereas GPAW yields the smallest value, 0.389~pm/V.
For $d_{24}$ and $d_{32}$, CASTEP and GPAW give similar values, whereas VASP yields a larger magnitude for $d_{24}$ and a smaller magnitude for $d_{32}$.
Accordingly, CASTEP gives near equality for $(d_{15},d_{31})$, with $\Delta=0.7\%$, and a modest mismatch for $(d_{24},d_{32})$, with $\Delta=3.42\%$.
GPAW yields $\Delta=4.28\%$ and $3.58\%$ for the same pairs.
VASP yields larger deviations, with $\Delta=11.69\%$ and $13.44\%$.

The results obtained with $\epsilon=10^{-7}$~Ha are similar to those obtained with $\epsilon=10^{-4}$~Ha for F-KDP and BNPO.
For F-KDP, $\Delta(d_{15},d_{31})$ decreases from $4.03\%$ to $3.68\%$ in the GPAW data when $\epsilon$ is reduced from $10^{-4}$ to $10^{-7}$.
For P-KDP, $\Delta(d_{14},d_{36})$ remains within 1\% for all three codes.

In summary, Table~\ref{tbl:shg} shows that the tensor pattern and dominant magnitudes are broadly reproducible across electronic-structure backends, whereas the internal consistency between Kleinman-symmetry-related pairs, as quantified by $\Delta$, can vary appreciably across implementations.
The larger sensitivity observed for both F-KDP and the independent BNPO test case, which share point group $mm2$, relative to P-KDP, which belongs to point group $\overline{4}2m$, is consistent with the hypothesis that lower crystallographic symmetry increases numerical sensitivity.
However, Table~\ref{tbl:shg} provides only a comparison at fixed computational settings.
It therefore cannot distinguish an intrinsic symmetry dependence from incomplete convergence or package-specific approximations.
We next examine how the same symmetry metrics evolve with the size of the conduction-band space and identify which part of the response controls the residual mismatch.

\InputIfFileExists{ext_shg_tbl.itex}{}{\message{LaTeX Warning: File `\filename@base.\ifx\filename@ext\relax tex\else\filename@ext\fi' not found on input line \the\inputlineno}

\subsection{Conduction-band convergence and numerical origin of the symmetry mismatch}

The fixed-setting comparison above suggests that lower symmetry is associated with increased numerical sensitivity, but it also reveals substantial differences among CASTEP, GPAW, and VASP.
To separate these effects, we now track the Kleinman-symmetry metrics as functions of the number of included conduction bands, $N_c$.
As discussed in Sec.~\ref{sec:methods}, this test directly probes the finite-band approximation used to evaluate the generalized derivatives.

Figures~\ref{fig:pkdp_prop_vs_Nc}--\ref{fig:bnpo_prop_vs_Nc} show the convergence of the symmetry-related component pairs with increasing $N_c$.
For P-KDP, Fig.~\ref{fig:pkdp_prop_vs_Nc}(a) plots $\Delta(d_{14},d_{36})$ computed with CASTEP, GPAW, and VASP using two energy-denominator regularization parameters, $\epsilon=10^{-4}$ and $10^{-7}$~Ha.
Figure~\ref{fig:pkdp_prop_vs_Nc}(b) shows the corresponding absolute difference $d_{14}-d_{36}$, in pm/V, together with the decomposition into contributions from $\chi^{abc}$, $\chi^{abc}_i$, and $\chi^{abc}_{i,\mathrm{two}}$ for $\epsilon=10^{-4}$~Ha.
The corresponding analyses for F-KDP and BNPO are shown in Figs.~\ref{fig:fkdp_prop_vs_Nc} and \ref{fig:bnpo_prop_vs_Nc}.
In those figures, panels (a) and (b) report $\Delta$ for the two symmetry-related pairs, whereas panels (c) and (d) report the corresponding absolute differences and their decompositions.
Results for $\epsilon=10^{-7}$~Ha are provided in Figs.~S7--S9 for P-KDP, F-KDP, and BNPO, respectively.
Those figures also show the convergence of the averaged values $(d_{ij}+d_{kl})/2$ with $N_c$.

For all three materials, the $\chi^{abc}_i$ contribution closely follows the total $\chi^{abc}$ contribution in the $\Delta$ panels.
This indicates that the mismatch between symmetry-related components is dominated by $\chi^{abc}_i$ rather than by $\chi^{abc}_e$.
This trend is consistent with the discussion in Sec.~\ref{sec:methods}.
In addition, the contribution from $\chi^{abc}_{i,\mathrm{two}}$ is nearly independent of $N_c$.
Therefore, the $N_c$ dependence of $\Delta(d_{ij},d_{kl})$ primarily arises from $\chi^{abc}_{i,\mathrm{three}}$.
Within the Kleinman-symmetrized static formulation, the two-band term $\chi^{abc}_{\mathrm{II}}$ [Eq.~\eqref{eq:chi_II}] vanishes under TRS.
Thus, $\chi^{abc}_{i,\mathrm{two}}$ should not be interpreted as an independent physical contribution in the static limit.
Instead, its two-band-like contribution is compensated by the corresponding part of $\chi^{abc}_{i,\mathrm{three}}$ when the intermediate-state sum is complete.
Incomplete compensation caused by conduction-band truncation can therefore lead to residual differences between Kleinman-symmetry-related components.

For P-KDP, $\Delta(d_{14},d_{36})$ remains below $2\%$ for all codes and for both $\epsilon$ values.
It approaches the percent level once $N_c \gtrsim 400$ and then exhibits only weak fluctuations as $N_c$ is increased further.
Consistently, Fig.~\ref{fig:pkdp_prop_vs_Nc}(b) shows that $d_{14}-d_{36}$ is extremely small, below $10^{-2}$~pm/V.
This difference is orders of magnitude smaller than the individual coefficients.
The residual $\Delta$ therefore reflects a minute imbalance between the two symmetry-related components rather than a qualitative breakdown of the tensor relations.
This behavior is consistent with Ref.~\citenum{Jia2020}.

For F-KDP and BNPO, the convergence is more code dependent and more sensitive to $\epsilon$.
At large $N_c$, CASTEP and GPAW typically reduce the mismatches to the few-percent level, and the corresponding absolute differences stabilize at a few $\times 10^{-2}$~pm/V.
In CASTEP, however, $\Delta(d_{15},d_{31})$ and $\Delta(d_{24},d_{32})$ decrease as $N_c$ increases from 200 to about 400 and then increase monotonically with further increasing $N_c$.
By contrast, the GPAW and VASP results show a more conventional saturation behavior.
Additional numerical checks with denser $k$-point meshes and tighter self-consistent convergence criteria do not remove this trend.
A plausible origin is the velocity-operator implementation in CASTEP.
Optical matrix elements are evaluated via $[\hat{H},\mathbf{r}]=\hat{v}$, which does not necessarily reduce to $\hat{p}/m_e$ in the presence of nonlocal potentials.
As a result, terms involving $\partial_{k_b}\partial_{k_a}\hat H$ can contribute to the generalized derivatives in Eq.~\eqref{eq:gen_deriv_sum_rule}.

The VASP data show larger and more persistent mismatches for the same pairs, particularly for F-KDP, and they also yield larger absolute differences.
This indicates that the residual asymmetry is not solely due to conduction-band truncation, but is also sensitive to implementation details and post-processing conventions.
A similar tendency was noted in Ref.~\citenum{Jia2020}, where reduced symmetry was associated with stronger apparent violations of Kleinman relations.
For F-KDP, the $\chi_{i,\mathrm{two}}$ contribution to $d_{15}-d_{31}$ obtained with VASP is comparable to those obtained with CASTEP and GPAW [Fig.~\ref{fig:fkdp_prop_vs_Nc}(c)], whereas this agreement does not hold for $d_{24}-d_{32}$.
This again points to package-dependent behavior of $\chi_{i,\mathrm{three}}$ for certain component pairs.
Overall, these figures show that increasing $N_c$ reduces the apparent symmetry breaking associated with incomplete conduction-band summations, while the large-$N_c$ mismatch can remain method dependent for the more sensitive cases of F-KDP and BNPO.

\begin{figure}[tbp]
    \centering
    \includegraphics[scale=1.0]{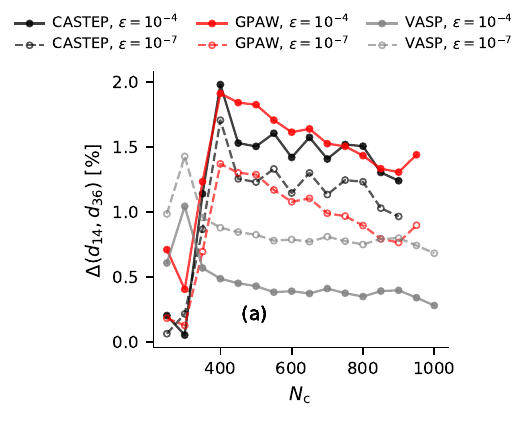}\\
    \includegraphics[scale=1.0]{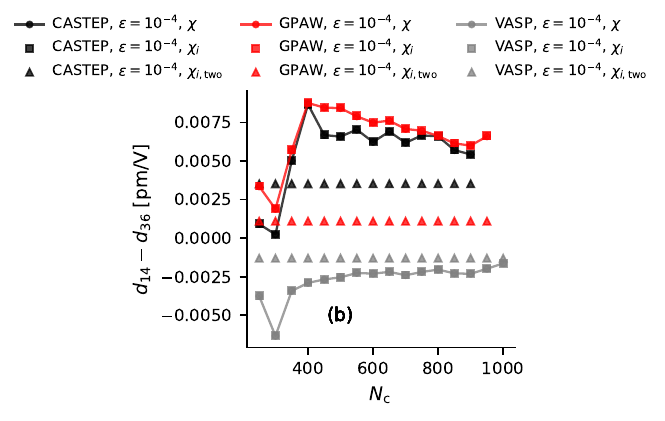}
    \caption{
    Conduction-band convergence of the P-KDP symmetry diagnostics.
    Panel (a) shows $\Delta(d_{14},d_{36})$ in \%, and panel (b) shows the absolute difference $d_{14}-d_{36}$ in pm/V, as functions of the number of included conduction bands $N_c$.
    }
    \label{fig:pkdp_prop_vs_Nc}
\end{figure}

\begin{figure}[tbp]
    \centering
    \includegraphics[scale=1.0]{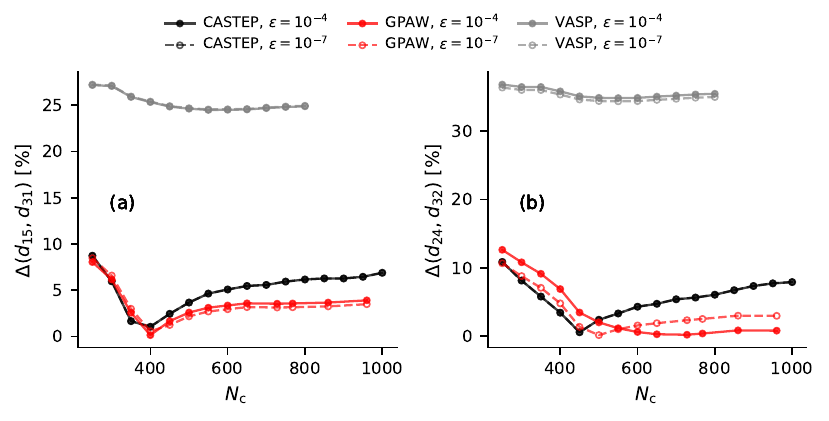}\\
    \includegraphics[scale=1.0]{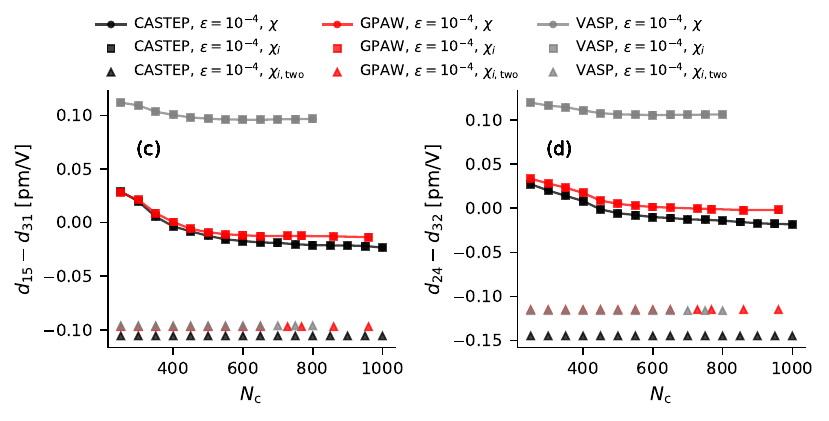}
    \caption{
    Conduction-band convergence for F-KDP.
    Panels (a) and (b) show $\Delta(d_{15},d_{31})$ and $\Delta(d_{24},d_{32})$ in \%, respectively.
    Panels (c) and (d) show the corresponding absolute differences $d_{15}-d_{31}$ and $d_{24}-d_{32}$ in pm/V as functions of $N_c$.
    }
    \label{fig:fkdp_prop_vs_Nc}
\end{figure}

\begin{figure}[tbp]
    \centering
    \includegraphics[scale=1.0]{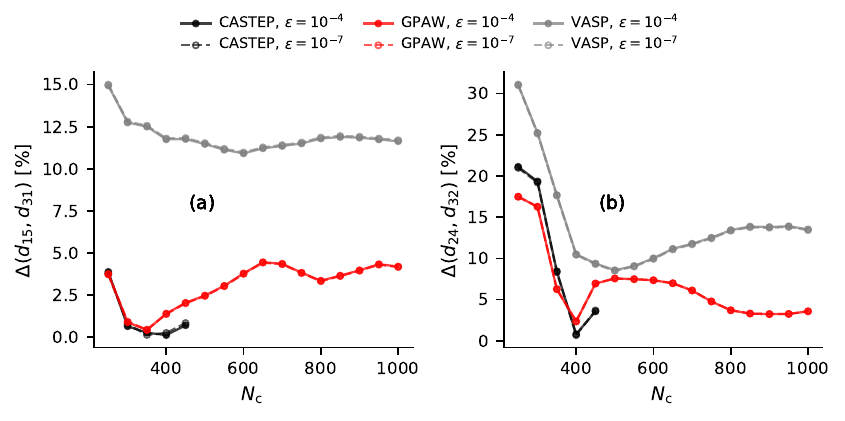}\\
    \includegraphics[scale=1.0]{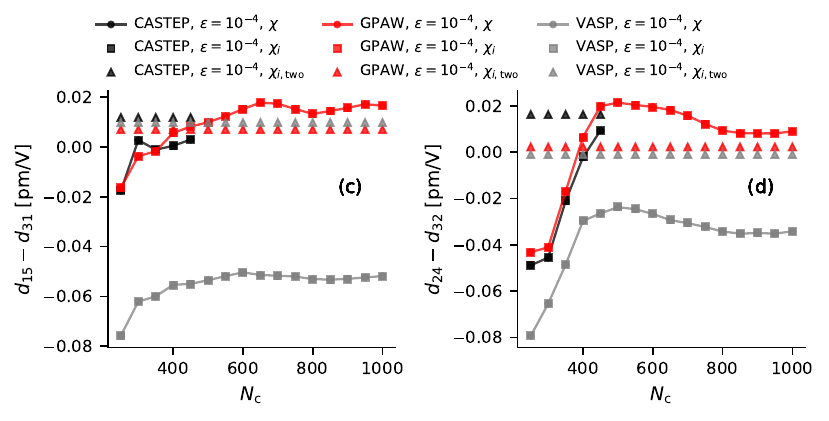}
    \caption{
    Conduction-band convergence for BNPO.
    Panels (a) and (b) show $\Delta(d_{15},d_{31})$ and $\Delta(d_{24},d_{32})$ in \%, respectively.
    Panels (c) and (d) show the corresponding absolute differences $d_{15}-d_{31}$ and $d_{24}-d_{32}$ in pm/V as functions of $N_c$.
    }
    \label{fig:bnpo_prop_vs_Nc}
\end{figure}

    \section{Summary}
    \label{sec:summary}
    We investigated how two scissors-correction schemes affect first-principles predictions of SHG in representative UV/DUV-NLO crystals, focusing on the widely used scheme-L and scheme-N.
Although these two prescriptions were compared when scheme-N was introduced, scheme-L remains the default in several electronic-structure packages and, for some tensor components and experimental datasets, can yield closer agreement with experiment.
To enable a consistent and numerically robust assessment, we derived a unified static-limit formulation that avoids numerical divergences and can be applied to both scheme-L and scheme-N, thereby extending earlier treatments that were effectively limited to scheme-L.
We implemented these developments in our Python package \texttt{NLOkit}, which provides a common interface for SHG calculations across multiple first-principles backends and enables controlled cross-code comparisons.

Using \texttt{NLOkit}, we performed systematic tests for representative borate and phosphate UV/DUV-NLO crystals and quantified scheme-dependent trends in both static coefficients and frequency-dependent spectra.
Across all compounds examined, the two scissors-correction schemes largely preserve the spectral line shape while primarily rescaling the overall magnitude, with scheme-N systematically yielding larger SHG responses than scheme-L.
For the largest tensor component of each crystal included in the static-limit comparison, the enhancement produced by scheme-N is about 15\%--25\%.
Comparison with experiment at 1064~nm further shows that the choice of scissors-correction scheme leads to a systematic shift in SHG magnitude, but the question of which scheme is closer to experiment remains influenced by the non-negligible spread among available measurements.
In the static limit, Kleinman symmetry is satisfied at the level of the formal expressions, whereas the apparent symmetry breaking found in practical calculations arises mainly from the numerical approximation used to evaluate generalized derivatives.
This numerical sensitivity is weak for P-KDP, but can become appreciable for the lower-symmetry F-KDP and BNPO cases, which share point group $mm2$.
The inclusion of BNPO as an independent test supports a role for lower crystallographic symmetry, while the different cross-code convergence trends show that the residual mismatch remains implementation dependent.
Overall, our results clarify the quantitative impact of scissors-correction-scheme choice on SHG predictions in representative UV/DUV-NLO crystals, provide a numerically stable formalism applicable to both scheme-L and scheme-N, and establish \texttt{NLOkit} as a practical platform for reproducible SHG calculations and diagnostics across \textit{ab initio} codes.

    \begin{acknowledgments}
        This work is supported by the Key Research Program of Frontier Sciences, National Natural Science Foundation of China (22193044, 52403305, 22361132544), CAS Project for Young Scientists in Basic Research (YSBR-024), the Strategic Priority Research Program of the Chinese Academy of Sciences (XDB0880000).

    \end{acknowledgments}

    \section*{Data availability}
        All data supporting the findings of this study are included in the manuscript and the supplementary material.

    \section*{Supplementary Material}
        See the supplementary material for the complete length-gauge SHG expressions and static-limit derivations; the calculated linear dielectric tensors and principal refractive indices at 1064~nm; additional frequency-dependent SHG magnitude spectra and representative imaginary-part spectra; and additional cross-code static SHG data and conduction-band-convergence analyses for P-KDP, F-KDP, and BNPO.

    \bibliography{references}

\end{document}